\documentclass[aps,prl,twocolumn,groupedaddress,showpacs,preprintnumbers,amsmath,amssymb]{revtex4}

\usepackage{graphicx}
\usepackage{dcolumn}
\usepackage{bm}

\begin{document}


\title{Investigation of Heterodiffusion in Molecular Crystals by the Method
of Raman Effect}



\author{M.A. Korshunov}
\affiliation{L.V. Kirensky Institute of Physics, Siberian Branch of Russian Academy of Sciences, Krasnoyarsk, 660036, Russia}
\email[]{mkor@iph.krasn.ru}

\date{\today}

\begin{abstract}
Using a method of a Raman effect of light the
heterodiffusion in molecular crystals is investigated by the
example of p-dichlorobenzene in a p-dibromobenzene. The installation diagram
and a procedure of a determination of diffusivities using this method is
given. Changes of diffusivities in a cut of a crystal perpendicularly to a
direction of diffusion are found. Obtained values of diffusivities are in
good agreement with those obtained by other methods.
\end{abstract}


\maketitle



Perspectives of use of organic crystals of a low symmetry in the molecular
electronics engineering for recording and information processing are
mentioned \cite{ref1}. For reading recorded information, it is possible to use the laser. Molecular crystals have rather low melting point and such exposure of
crystals by a laser beam can lift temperature of a sample close to melting
point that will affect magnification of diffusion of molecules of impurity
and durability of recorded information. Therefore it is actual to
investigate diffusion in molecular crystals. As against processes of
diffusion in inorganic materials, diffusion in molecular crystals has the
singularities. The difficulties originating at investigation of migration of
molecules are in particular connected with nonsphericity of molecules,
presence of orientation oscillations, deficiencies, etc., that distinguishes
them from inorganic systems.

In order to investigate the process of diffusion various methods are used
(chemical, metallographic and physical \cite{ref2}) which majority at the
quantitative investigation of this process determines concentration of
impurity in a sample on different depth after its ageing at fixed
temperature for a long time. At that, the \textit{average} value of impurity concentration
throughout the entire plane of a crystal cut perpendicular to a direction of
diffusion is determined frequently. Similar methods were used and at
investigation of molecular crystals \cite{ref3}. More precise is the method the
local X-ray spectral analysis used for investigation of metals and alloys
\cite{ref4}.

The purpose of the present work is to consider a possibility of use of the
method of Raman Effect of light (RE) for investigation of impurity
heterodiffusion in a monocrystal, grown by Bridgman's method.

By the method of a Raman Effect it is possible to determine concentration of
impurity as along a direction of diffusion as perpendicular to this
direction, that enables to investigate a heterodiffusion in molecular
crystals. Moreover, this method allows to judge by spectrums of the lattice
and intramolecular oscillations about a character of a disposition of
impurity molecules among molecules of the base crystal. In particular, from
Raman spectra one can determine whether the solid solution is formed by the
substitution (i), or interstitial (ii), or the impurity molecules are
concentrated at the boundaries between the blocks of the single-crystal
lattice (iii) or the solid solution is the mechanical mixture of the
substances (iv).

\begin{figure}
\includegraphics[width=0.47\textwidth]{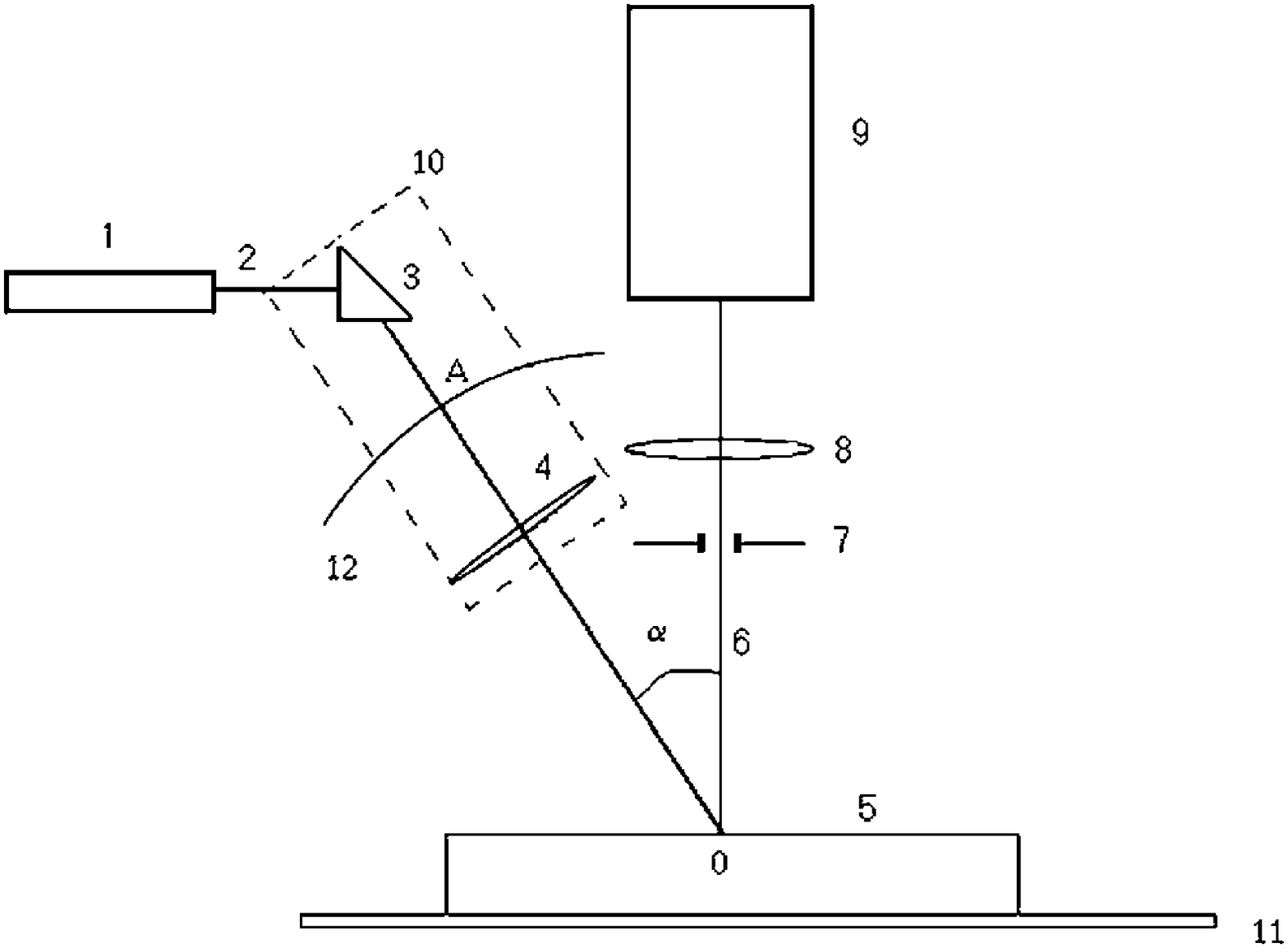}
\caption{\label{fig1} The installation diagram for investigation of diffusion in molecular crystals by the method of RE.}
\end{figure}

The monocrystal placed in installation which diagram is shown in figure~\ref{fig1}.
This installation allows to obtain a spectrum of the RE from area on a
surface of a crystal irradiated with the focused laser beam.

The installation diagram in figure~\ref{fig1} consist of He-Ne laser (1) which ray
(2), transited through a prism, that rotates it (3), and a lens (4), then
focalized on a surface of investigated sample (5). Radiation of a crystal
(6), transited through a diaphragm (7) and a lens (8), hits on a slit of a
spectrograph (9). By changing incidence angle $\alpha $ of a laser beam on a
sample in a point $O$ the best ratio signal/noise was found. Rotational on
radius $OA$ was carried out by a system lens-prism (10). The sample was on a
table (11) which was moved by the micrometer screw that allowed to carry out
recording of spectrums while moving a sample on small distances. The
diaphragm (7) is used for a cut of a laser ray reflected from a point $O$.

Because the diameter of a stain of laser beam can be reduced on the surfaces
of a crystal up to $2\,\mu $ ($\mu $ stands for micrometer) this method
allows to investigate concentration of components along a direction of
diffusion.

As subjects of inquiry the substances well investigated by various methods
have been chosen. As the base monocrystal the p-dibromobenzene and as
diffusing impurity the p-dichlorobenzene has been chosen. A p-dibromobenzene
(measured melting point $t_m = 86.8\pm 0.10\,^{\circ}C)$ and
p-dichlorobenzene ($t_m = 53.0\pm 0.10\,^{\circ}C)$. P-dichlorobenzene
($\alpha $-modification) and isomorphous to it p-dibromobenzene are
investigated by optical, radiographic and NQR methods \cite{ref5,ref6,ref7}. Both substances
are crystallizes in space group $P2_1 / a$ with two molecules in a unit cell
and forms a mixed crystals of substitution at any concentrations of
components (this could be determined as by the phase diagram, as by
spectrums of the RE \cite{ref6}). Spectrums of the lattice oscillations are similar
to spectrums of components. Monocrystals are pellucid, that allows obtaining
Raman spectrums of high quality.

Before investigation of diffusion of p-dichlorobenzene in a
p-dibromobenzene, the monocrystals of solid solutions of investigated
substances have been grown and spectrums of a Raman Effect are obtained in
the range of the lattice and intramolecular oscillations (up to $400\,cm^{ -
1})$. In present paper the concentration of components was measured in molar
unities.

First, spectrums of standard samples with the given relation $p - C_6 H_4
Br_2 $ and $p - C_6 H_4 Cl_2 $ have been obtained. There is well defined
line at frequency $\nu = 20.0\,cm^{ - 1}$ in a spectrum of the lattice
oscillations of a p-dibromobenzene. In a spectrum of a mixed crystal the
frequency $\nu $ monotonically varies depending on concentration of
components. This dependence is presented in work \cite{ref8}. This allows
determining of concentration of impurity (p-dichlorobenzene) in investigated
solid solutions by the value of frequency. The value of frequency was
determined with accuracy $\pm 0.2\,cm^{ - 1}$.

\begin{figure}
\includegraphics[width=0.47\textwidth]{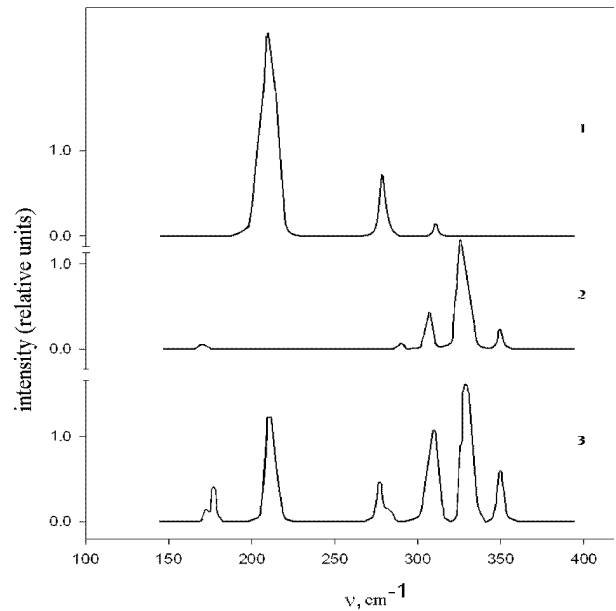}
\caption{\label{fig2} Intramolecular oscillations spectrums of $p - C_6 H_4 Br_2 $ (1) and $p - C_6 H_4 Cl_2 $ (2) and mixed crystal, which has been grown from these substances, at equimolar concentration (3), in range of frequencies $\nu = 150 - 400\,cm^{ - 1}$.}
\end{figure}

If in a spectrum of the lattice oscillations there is separately located and
well defined line then in order to determine concentration of impurity it is
possible to use a method described above. However, lines in a spectrum are
frequently so disposed that it is very difficult to isolate the separately
located line. Then the magnitude of concentration of impurity can be
determined on relative intensity of the valent intramolecular oscillations.
According to work \cite{ref9}, the line in a spectrum of the RE $p - C_6 H_4 Br_2 $
with $\nu = 212.0\,cm^{ - 1}$ matches to valence vibration of $C - Br$, and
a line with $\nu = 327.0\,cm^{ - 1}$ in $p - C_6 H_4 Cl_2 $ matches to
valence vibration of $C - Cl$. In figure~\ref{fig2} spectrums of the intramolecular
oscillations (in range of frequencies $\nu = 150 - 400\,cm^{ - 1})$ of
p-dibromobenzene (1), p-dichlorobenzene (2) and an investigated mixed
crystal at $50\,mol\% $ of p-dichlorobenzene (3) are presented.

\begin{figure}
\includegraphics[width=0.47\textwidth]{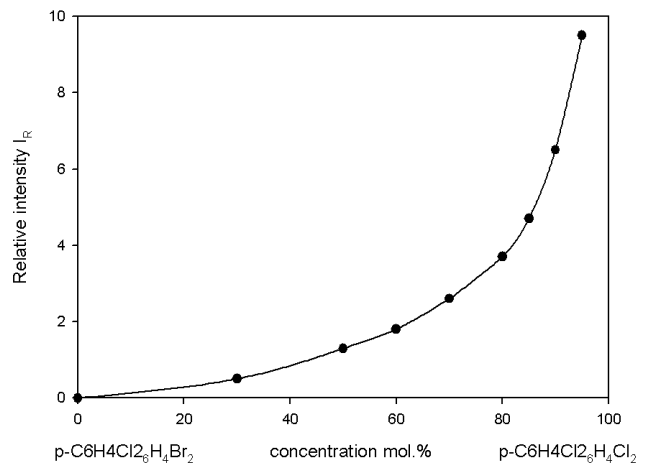}
\caption{\label{fig3} Dependence of relative intensity of a line with frequency of $327\,cm^{ - 1}$ to a line with frequency of $212\,cm^{ - 1}$ on concentration of $p - C_6 H_4 Cl_2 $ in investigated mix-crystals.}
\end{figure}

Apparently, the spectrum of a mixed crystal is a superposition of spectrums
of the intramolecular oscillations of components with account for their
concentration. In figure~\ref{fig3} the graph of dependence of relative intensity of
a line with frequency $327\,cm^{ - 1}$ to a line with frequency $212\,cm^{ -
1}$ from concentration of components is presented. With use of these graphs
it was possible to determine concentration of p-dichlorobenzene in a
p-dibromobenzene. In greater detail the method of determination of
components concentration in a mixed crystal with use of the RE is described
in work \cite{ref8}.

Then the monocrystal of a p-dibromobenzene by the Bridgman's method has been
grown. The monocrystal was grown in a glass tube with a diameter $d = 1\,cm$
and with a length $h = 1\,cm$ with the drawn end in which seated beforehand
purified by a method of zone melting a p-dibromobenzene. From a tube the air
was pumped out and the tube was soldered. The ampoule with substance was
sinking from the hot to a cold zone of the heater with velocity $V = 8.3
\cdot 10^{ - 6}\,cm / s$. The lapse rate of temperature of the heater was
set by various coiling of a heating coil and was $dt / dl = 7.6\,\deg  /
cm$. Monocrystallinity of a sample was tested with use of a polarized-light
microscope. The parallelepiped was cut out from this monocrystal ($0.5\times
0.5\times 7.0\,cm^3)$ and on one of which least edges the layer of
p-dichlorobenzene with thickness $0.03\,cm$ was applied. After that the
sample was annealed at temperature $50^{\circ}C$ during 360 hours.

To exclude surface effects and to investigate diffusion in depth of a sample
from one of the edges the layer of a sample by thickness $0.1\,cm$
collateral to a direction of diffusion has been cut off. The sample was put
in installation, which diagram is presented in figure~\ref{fig1}, and the recording
of spectrums was carried out. Spectrums are obtained in three points located
at $20\,\mu $ from boundary parting p-dichlorobenzene and a p-dibromobenzene
along a direction of diffusion. At that the concentration $C$ of
p-dichlorobenzene (determined using method described above) is $C_{20} =
35.0\,mol\% $, $C_{30} = 16.0\,mol\% $ and $C_{40} = 40.0\,mol\% $, where
lower index designates distance in $\mu $ from boundary. Apparently, the
used monocrystal can be considered as semi-infinite body (in comparison with
penetration depth of impurity during diffusion) and consequently use the
following relation for a determination of a diffusivity $C = C_0 erfcZ$,
where $Z = x / 2\sqrt {Dt} $ \cite{ref10}, $C$ is the concentration of impurity
located at $x$ from boundary, $C_0 $ is the initial concentration of
impurity, $t$ is the time of annealing, $D$ is the diffusivity.

It is found that the diffusivity is $D_{20} = 1.76\pm 0.02 \cdot 10^{ -
12}\,cm^2 / s$, $D_{30} = 1.75\pm 0.02 \cdot 10^{ - 12}\,cm^2 / s$, $D_{40}
= 1.73\pm 0.02 \cdot 10^{ - 12}\,cm^2 / s$ and average value is $D =
1.746\pm 0.02 \cdot 10^{ - 12}\,cm^2 / s$.

\begin{figure}
\includegraphics[width=0.47\textwidth]{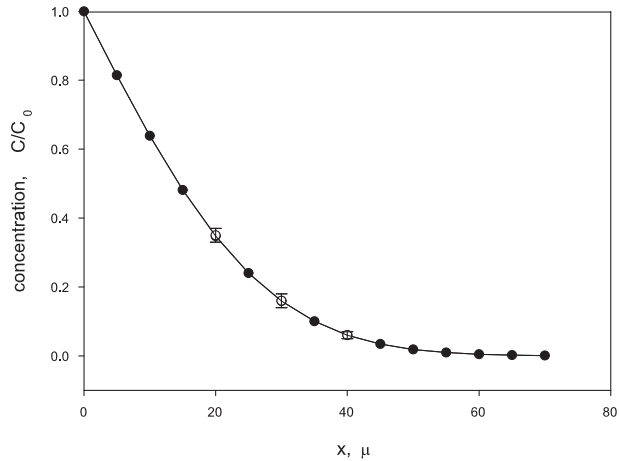}
\caption{\label{fig4} Dependence of relative concentration $C / C_0 $ of $p - C_5 H_4 Cl_2 $ on depth of penetration in the $p - C_5 H_4 Br_2 $ monocrystal. Circles ($\circ$) mark the experimental values of concentration.}
\end{figure}

In figure~\ref{fig4} the calculated dependence of p-dichlorobenzene concentration on
depth of penetration in a monocrystal of a p-dibromobenzene for the obtained
average value of diffusivity $D$ is presented. Concentration was calculated
with use of above-stated formulas. In the figure the circles ($\circ$) marks the experimental averaged values of concentration. Evidently,
experimental values are in very good agreement with calculated
concentrations.

Obtained values of diffusivity are in good agreement with the data obtained
on other molecular crystals by other methods \cite{ref11,ref12}. In work \cite{ref11} it is
mentioned, that magnitude of a diffusivity strongly depends on a degree of
clearing of substance, for example for naphthalene, from $10^{ - 10}$ to
$10^{ - 14}\,cm^2 / s$.

\begin{figure}
\includegraphics[width=0.47\textwidth]{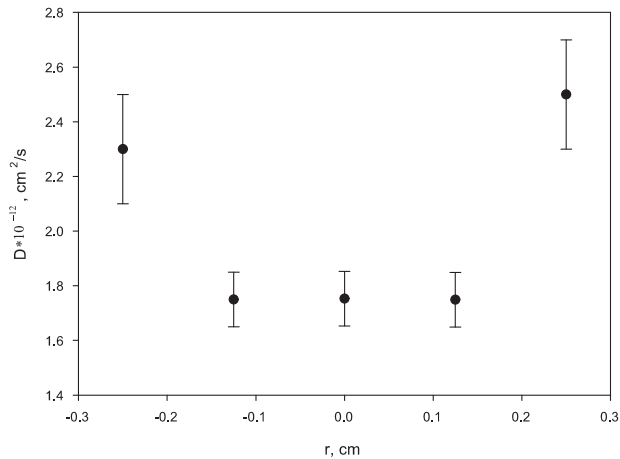}
\caption{\label{fig5} Changes of diffusivity on centerline of obtained sample perpendicularly to a direction of diffusion at $20\,\mu $ from boundary that parting p-dichlorobenzene and a p-dibromobenzene.}
\end{figure}

Then, using procedure given above, the diffusivity in a sample
perpendicularly to a direction of diffusion at about $20\,\mu $ from
boundary was measured. With this purpose on this distance the collateral
layer of a crystal perpendicularly to a direction of diffusion has been cut
off. Spectrums are obtained in five points on centerline of this sample
perpendicularly to a direction of diffusion. The values of diffusivities
found from experiments are marked by circles in figure~\ref{fig5}. Evidently, to
edges of a crystal these values increase.

Thus, using a method of a Raman Effect of light and the above-stated
procedure it is possible to carry out investigations of a heterodiffusion in
molecular crystals. Values of diffusivities obtained in this paper are in
good agreement with given obtained by other methods on molecular crystals.


\end{document}